# AN IMPROVED LOWER BOUND OF THE SPARK WITH APPLICATION

Wajeb Gharibi

College of Computer Science and Information Systems
Jazan University, Kingdom of Saudi Arabia
gharibi@jazanu.edu.sa

## *ABSTRACT*

*Spark plays a great role in studying uniqueness of sparse solutions of the underdetermined linear equations. In this article, we derive a new lower bound of spark. As an application, we obtain a new criterion for the uniqueness of sparse solutions of linear equations.*

## 1. Introduction

Recent theoretical developments have generated a great deal of interest in sparse signal representation. The setup assumes a given dictionary of "elementary" signals, and models an input signal as a linear combination of dictionary elements, with the provision that the representation is sparse, i.e., involves only a few of the dictionary elements. Finding sparse representations ultimately requires solving for the sparsest solution of an underdetermined system of linear equations. Such models arise often in signal processing, image processing, and digital communications.

Given an $A \in R^{n \times m}$ $(n < m)$ full-rank matrix with no zero columns $b \in R^n$, the linear system $Ax = b$ has infinitely many solutions when the system is underdetermined. Depending on the nature of source problems, we are often interested in finding a particular solution, and thus optimization methods come into a play through certain merit functions that measure the desired special structure of the solution. One of the recent interests is to find the sparsest solution of an underdetermined linear system, which has found many applications in signal and image processing [1-3]. To find a sparsest solution of $Ax = b$, perhaps the ideal merit function is the cardinality of a vector, denoted by $\|x\|_0$ i.e., the number of nonzero components of $x$. Clearly, the set of the sparsest solutions of $Ax = b$ coincides with the set of solutions to the cardinality minimization problem:

$$\min\{\|x\|_0 : Ax = b\}, \qquad (1.1)$$

which is an NP-hard discrete optimization problem [4]. The recent study in the field of compressed sensing nevertheless shows that not all cardinality minimization problems are equally hard, and there does exist a class of matrices A such that the problem (1.1) is computationally tractable. These matrices can be characterized by such concepts as the spark which was formally defined by Donoho and Elad [5], restricted isometry property (RIP) introduced by Cand`es and Tao [6], mutual coherence (MC) [7-9], and null space property (NSP) [9, 12].

Problem (1.1) is not easy to solve in general. From a convex analysis point of view, a natural methodology is to minimize the convex envelope of $\|x\|_0$. It is well-known that $\ell_1-norm$ is the convex envelope of $\|x\|_0$ over the region $\{x: \|x\|_\infty \leq 1\}$. One of the main approaches to attack (1.1) is through $\ell_1-\min imization$

$$\min\{\|x\|_1: \ Ax=b\}, \tag{1.2}$$

which is identical to a linear program (LP) and hence can be solved very efficiently. Using $\ell_1-norm$ as a merit function for sparsity can be traced back several decades in a wide range of areas from seismic traces [13], sparse-signal recovery[14], sparse-model selection (LASSO algorithm) in statistics [15] to image processing [16], and continues its growth in other areas [17-19].

Let $A \in R^{n \times m}$ ($n < m$) be any full-rank matrix with no zero columns. The definition of spark was first introduced by Donoho and Elad [5] as follow:

**Definition 1.1** ([3, 5]). The spark of a given matrix $A$, denoted by $spark(A)$, is the smallest number of columns from $A$ that are linearly dependent.

Clearly, $1 \leq spark(A) \leq n+1$. In general, it is difficult to calculate the spark. Actually, to obtain it, a combinatorial search over all possible subsets of columns from $A$ is required.

Nevertheless, the spark can be easily lower bounded by using the term 'mutual coherence'.

**Definition 1.2** ([3, 5, 7]). The mutual coherence of a given matrix $A$ is the largest absolute normalized inner product between different columns from $A$. Denoting the k-th column in $A$ by $a_k$, the mutual coherence is given by

$$\mu(A) = \max_{1 \leq k < j \leq m} \frac{|a_k^T a_j|}{\|a_k\|_2 \|a_j\|_2} \tag{1.3}$$

Notice that $\mu(A) > 0$ under the assumption $n < m$.

**Theorem 1.1** ([3, 5]). For any matrix $A \in R^{n \times m}$ ($n < m$) with no zero columns, the following relationship holds:

$$spark(A) \geq 1 + \frac{1}{\mu(A)} \tag{1.4}$$

In this article, we first derive a new lower bound of $spark(A)$ as an improvement of Theorem 1.1, illustrated by an example. As a direct application, we obtain a new criterion for uniqueness of sparse solutions of linear equations. Finally, we extend the definition of spark and the related result for the general matrix $A$.

The rest of our paper is structured as follows: In Section 2, we give the motivation of our discussed problem. In Section 3, we develop a new lower bound for the sparse supported with theorems and example. In Section 4, we set an application by considering the problem of finding the sparse solutions of linear equations. In Section 5, we extend the definition of the spark to the general matrix $A \in R^{n \times m}$ with no zero columns. Conclusion is given in Section 6.

## 2. Motivation: Signal and Image Compression

Generally, solving the sparse solution of underdetermined systems of linear equations has many applications in wide areas such as signal processing and image compression, Compressed Sensing, Error Correcting Codes, Recovery of Loss Data, and in Cryptography. Here, we explain the most direct and natural application; Signal and Image Compression.

Suppose that a signal or an image $b$ is represented by using a tight frame $A$ of size $m \times n$ with $m \leq n$. We look for a sparse approximation x satisfying

$$\min\{\|x\|_0 : x \in R^n, \|Ax - b\| \leq \theta\}, \tag{2.1}$$

where $\theta > 0$ is a tolerance. In particular, for lossless compression, i.e., $\theta = 0$, the above described problem (2.1) is the same problem (1.1).

## 3. A New Lower Bound

Denote by $a_k \neq 0$ the k-th column in $A$. For each index pair $(k, j)$ $(1 \leq k < j \leq m)$, define

$$\mu_{kj} = \frac{|a_k^T a_j|}{\|a_k\|_2 \|a_j\|_2} \tag{3.1}$$

Where $\|.\|_2$ is the standard $l_2-norm$. Rearrange $\mu_{k,j}$ by sorting them in non-increasing order:

$$\mu_1 \geq \mu_2 \geq \cdots \geq \mu_{\frac{m(m-1)}{2}} \tag{3.2}$$

**Lemma 3.1.** For any matrix $A \in R^{n \times m}$ $(n < m)$ with no zero columns, it holds that

$$\sum_{i=1}^{n} \mu_i \geq 1.$$

**Proof:** First we modify the matrix $A$ by normalizing its columns to unit $l_2-norm$, obtaining $\tilde{A}$. This operation preserves all $\mu_i$. The entries of the resulting Gram matrix $G = \tilde{A}^T \tilde{A}$ satisfy the following properties:

$$\begin{aligned} &\{G_{k,k} = 1 : 1 \leq k \leq m\}, \\ &\{|G_{k,j}| = \mu_{k,j} : 1 \leq k < j \leq m\}, \\ &\{|G_{k,j}| = \mu_{j,k} : 1 \leq j < k \leq m\}. \end{aligned} \tag{3.3}$$

We also have that $rank(G) \leq n$.

Consider $G_{n+1}$, an arbitrary minor from $G$ of size $(n+1) \times (n+1)$, built by choosing a subgroup of $n+1$ columns from $\tilde{A}$ and computing their sub-gram matrix.

Suppose $\sum_{i=1}^{n} \mu_i < 1$, then for every i, we have

$$\sum_{j \neq i} |(G_{n+1})_{i,j}| \leq \sum_{i=1}^{n} \mu_i < 1 = (G_{n+1})_{i,i},$$

i.e., $G_{n+1}$ is diagonally dominant according to the Gershgorin disk theorem [19], $G_{n+1}$ is positive definite and hence $rank\,(G_{n+1}) = n+1$, which contradicts the fact that $rank\,(G_{n+1}) \leq rank(G) \leq n$.

**Definition 3.1.** For any matrix $A \in R^{n \times m}$ $(n < m)$ with no zero columns, define the coherence index of $A$ as the minimum index $p$ such that $\sum_{i=1}^{p} \mu_i \geq 1$, denoted by $\gamma(A)$.

According to Lemma 2.1, the coherence index, γ(A), is well defined. Furthermore, one can write the formulation as follows:

$\gamma(A) = \min\ p$ (3.4)

$s.t.\ \sum_{i=1}^{p} \mu_i \geq 1,$ (3.5)

$p \in \{1,2,\dots,n\}.$ (3.6)

**Theorem 3.1**. For any matrix $A \in R^{n \times m}$ $(n < m)$ with no zero columns, it holds that

$$spark(A) \geq 1 + \gamma(A) \geq \frac{1}{\mu(A)}. \tag{3.7}$$

Proof. Similarly to the proof of Lemma 2.1, we normalize $A$ to $\tilde{A}$ which preserves both the spark and all $\mu i$. Let $G = \tilde{A}^T\tilde{A}$ whose entries satisfy the properties (2.3), as stated above.

Consider $G_P$, an arbitrary minor from $G$ of size $p \times p$ $(p \in \{2,3,\cdots,n+1\})$, built by choosing a subgroup of $p$ columns from $\tilde{A}$ and computing their sub-Gram matrix. From the Gershgorin disk theorem [1], if this minor $G_P$ is diagonally dominant, i.e. $\sum_{j \neq i} |(G_P)_{i,j}| < (G_P)_{i,i}$ for every $i$, then $G_P$ is positive definite, and so those $p$ columns from $\tilde{A}$ are linearly independent. The condition $\sum_{i=1}^{p-1} \mu_i < 1$ implies positive definiteness of every $p \times p\ G_P$. Thus, the condition $\sum_{i=1}^{p-1} \mu_i \geq 1$ is necessary for $G_P$ to be singular. (By Lemma 2.1, this condition is surely satisfied when $p = n+1$.) Therefore, $spark(A)$ is greater than or equal to the minimal $p$ such that $\sum_{i=1}^{p-1} \mu_i \geq 1$ i.e. $spark(A) \geq \min_{\sum_{i=1}^{p-1} \mu_i \geq 1} p = 1 + \gamma(A)$. By the definition of $\mu_i$, we notice that

$$\sum_{i=1}^{p} \mu_i \leq p\mu_1,$$

Then we have

$$\gamma(A) = \min_{\sum_{i=1}^{p} \mu_i^{\geq 1}} p \geq \min_{p\mu_1^{\geq 1}} p \qquad (3.8)$$
$$= \frac{1}{\mu_1} = 1/(\mu(A))$$

The proof is completed.

Below we use a trivial example to illustrate the possible improvement.

**Example 3.1.** Let $n \geq 2$ and

$$A = \begin{pmatrix} 1 & 0 & \cdots & 0 & 0.8 \\ 0 & 1 & \cdots & 0 & \frac{0.6}{\sqrt{n-1}} \\ \vdots & \vdots & \ddots & \vdots & \vdots \\ 0 & 0 & \cdots & 1 & \frac{0.6}{\sqrt{n-1}} \end{pmatrix}_{n \times (n+1)} \qquad (3.9)$$

Then
$$spark(A) = n + 1 \qquad (3.10)$$
$$1 + \gamma(A) = 2 + \left\lceil \frac{\sqrt{n-1}}{3} \right\rceil, \qquad (3.11)$$
$$1 + \frac{1}{\mu(A)} = 2.25, \qquad (3.12)$$

Where [x] returns the smallest integer value which is greater than or equal to x.

## 4. Application

In this section, we consider the problem finding the sparse solutions of linear equations, which is nowadays very popular, see [5] and references therein. The formulation of this problem can be written as follow:

$$\min \|x\|_0 \qquad (4.1)$$

$$s.t. \ Ax = b, \qquad (4.2)$$

Where $A \in R^{n \times m}$ and $\|x\|_0$ the so-called $l_0 - \ 'norm'$, denotes the number of nonnegative elements of x, i.e., $\|x\|_0 = \#\{i: x_i \neq 0\}$.

In general, Problem (4.1)-(4.2) is NP-hard [13, 14]. It is significant to study the uniqueness of the sparsest solution, which can be regarded as a sufficient condition for global optimality. The first surprising result is crucially based on the term 'spark'.

**Theorem 4.1** (Uniqueness-spark [5, 8]). If a system of linear equations $Ax = b$ has a solution $x$ obeying $\|x\|_0 < spark(A)/2$, this solution is necessarily the sparsest possible.

Notice that the spark is also difficult to obtain.

Following Theorem 1.1 and Theorem 4.1, one has

**Theorem 4.2** (Uniqueness-mutual coherence [5, 8]). If a system of linear equations $Ax = b$ has a solution $x$

obeying $\|x\|_0 < \frac{1}{2}\left(1 + \frac{1}{\mu(A)}\right)$, this solution is necessarily the sparsest possible.

Combining Theorem 3.1 with Theorem 4.1, we obtain the following result, which improves Theorem 4.2.

**Theorem 4.3** (Uniqueness-coherence index). If a system of linear equations $Ax = b$ has a solution $x$ obeying $\|x\|_0 < \frac{1}{2}(1 + \gamma(A))$, this solution is necessarily the sparsest possible.

## 5. Extension

In this section, we extend the definition of spark to the general matrix $A \in R^{n \times m}$ with no zero columns. The difference to the above assumption $n < m$ is that now it may happen that $rank(A) = m$. In this case, we set $spark(A) = +\infty$ as it may happen that $\mu(A) = 0$, otherwise, Theorem 1.1 does not make sense.

**Lemma 5.1**. Let $\mu_1, \ldots, \mu_{m-1}$ be defined in (2.2). If $\sum_{i=1}^{m-1} \mu_i < 1$ then $rank(A) = m$.

*Proof.* Similarly to the proof of Lemma 2.1, we normalize $A$ to $\tilde{A}$ which preserves both the spark and all $\mu_i$. Let $G = \tilde{A}^T\tilde{A}$ whose entries satisfy the properties (3.3), as stated above. Then for every $i$, we have

$$\sum_{j \neq i} |G_{i,j}| \leq \sum_{i=1}^{m-1} \mu_i < 1 = G_{i,i},$$

i.e., $G$ is diagonally dominant. According to the Gershgorin disk theorem [19], $G$ is positive definite and so the $m$ columns from $\tilde{A}$ are linearly independent.

**Theorem 5.1.** For any matrix $A \in R^{n \times m}$ with no zero columns, it holds that:

$$spark(A) \begin{cases} = +\infty, & \text{if } \sum_{i=1}^{m-1} \mu_i < 1, \\ \geq 1 + \gamma(A), & \text{otherwise,} \end{cases} \quad (5.1)$$

Where

$$\gamma(A) = \min p \quad (5.2)$$

$$s.t. \sum_{i=1}^{p} \mu_i \geq 1, \quad (5.3)$$

$$p \in \{1, 2, \ldots, m-1\}. \quad (5.4)$$

Proof. Following Lemma 3.1, we only need to consider the case $\sum_{i=1}^{m-1} \mu_i \geq 1$ similarly to the proof of Lemma 3.1, we normalize $A$ to $\tilde{A}$, which preserves both the spark and all $\mu_i$. Let $G = \tilde{A}^T\tilde{A}$ whose entries satisfy the properties (3.3), as stated above.

Consider an arbitrary minor from $G$ of size $p \times p$ ($p \in \{2, 3, \ldots, m\}$), built by choosing a subgroup of $p$ columns from $\widetilde{A}$ computing their sub-Gram matrix. From the Gershgorin disk theorem [19], if this minor is diagonally dominant, i.e. $\sum_{j \neq i} |G_{i,j}| < G_{i,i}$ for every $i$, then this sub-matrix of $G$ is positive definite, and so those $p$ columns from $\widetilde{A}$ are linearly independent.

The condition $\sum_{i=1}^{p-1} \mu_i < 1$ implies positive definiteness of every $p \times p$ minor, and so $\sum_{i=1}^{p-1} \mu_i \geq 1$ is a necessary condition for the $p \times p$ minor to be singular. Therefore, $spark(A)$ is greater than or equal to the minimal $p$ such that $\sum_{i=1}^{p-1} \mu_i \geq 1$ i.e.,

$$spark(A) \geq min_{\sum_{i=1}^{p-1} \mu_i \geq 1} p = 1 + \gamma(A).$$

Notice that the latter is well defined since $\sum_{i=1}^{m-1} \mu_i \geq 1.$

## 6. Conclusion

Non-negative linear systems of equations come up often in many applications in signal processing and image compression, compressed sensing, error correcting codes, recovery of loss data, to name just a few. Solving such systems is usually done by adding conditions such as maximal entropy, maximal sparsity, and so on.

In this work, we developed a new lower bound of spark supported by theorems and applications. We also have obtained a new criterion for the uniqueness of spark solutions of linear equations. In addition, we have extended the definition of the spark to a general matrix with no zero columns.

The obtained results generalize some of the previous- known facts about the spark and open the way to develop more potential stronger optimized bounds and solutions.

## Acknowledgement

I thank Prof. Yong Xia for his valuable comments that help improve the manuscript. I would also like to thank the referee for helpful comments.

## References

[1] D. Donoho, Compressed sensing, IEEE Trans. Inform. Theory, 52 (2006), pp. 1289-1306.

[2] E. Cand'es, Compressive sampling, International Congress of Mathematicians, Vol. III, 2006,

[3] A. Bruckstein, D. Donoho and M. Elad, From sparse solutions of systems of equations to sparse modeling of signals and images, SIAM Rev., 51 (2009), pp. 34-81.

[4] B. Natarajan, Sparse approximate solutions to linear systems, SIAM J. Comput., 24 (1995), pp. 227-234.

[5] Donoho and M. Elad, Optimality sparse representation in general (non-orthogonal) dictionaries via $\ell_1$-minimization, Proc. of the National Academy of Sciences, 100 (2003), pp. 2197-2202.

[6] E. Cand'es and T. Tao, Decoding by linear programming, IEEE Trans. Inform. Theory, 51(2005), pp. 4203-4215.

[7] S. Mallat and Z. Zhang, Matching pursuits with time-frequency dictionaries, IEEE Trans. Signal Process., 41 (1993), pp. 3397-3415.

[8] D. Donoho and X. Huo, Uncertainty principles and ideal atomic decomposition, IEEE Trans. Inform. Theory, 47 (1999), pp. 2845-2862.

[9]     R. Gribonval and M. Nielsen, Sparse decompositions in unions of bases, IEEE Trans. Inform. Theory, 49 (2003), pp. 3320-3325.

[10]    A. Cohen, W. Dahmen and R. DeVore, Compressed sensing and best k-term approximation, J. Amer. Math. Soc., 22 (2009), pp. 211-231.

[11]    Y. Zhang, Theory of compressive sensing via $\ell_1$- minimization: A Non-RIP analysis and extensions, Technical Report, Rice Univ., 2008.

[12]    H. Taylor, S. Banks and J. McCoy, Deconvolution with the $\ell_1$- norm, Geophysics, 44 (1979), pp. 39-52.

[13]    D. Donoho and P. Stark, Uncertainty principles and signal recovery, SIAM J. Appl. Math., 49(1989), pp. 906-931.

[14]    R. Tibshirani, Regression shrignkage and selection via the lasso, J. Royal Statist. Soc., B 58(1996), pp. 267-288.

[15]    P. Blomgren and T. F. Chan, Color TV: total variation methods for restoration of vector-valued images, IEEE Trans. Image Process., 7 (1998), pp. 304-309.

[16]    M. Fornasier and H. Rauhut, Compressive sensing, in the Handbook of Mathematical Methods in Imaging, Otmar Scherzer (ed.), Vol. 1, Springer, 2011.

[17]    I. F. Gorodnitsky and B. D. Rao, Sparse signal reconstruction from limited data using FO- CUSS: A re-weighted norm minimization algorithm, IEEE Trans. On Signal Processing, 45(3): (1997), pp. 600-616.

[18]    Sebastian Schmutzhard, Alexander Jung, Franz Hlawatsch, Zvika Ben-Haim, and Yonina C. Eldar. A Lower Bound on the Estimator Variance for the Sparse Linear Model, 44[th] Asilomar Conf. Signals, Systems, Computers, Pacific Grove, CA, Nov. 2010.

[19]    R. Bhatia, Matrix Analysis, New York: Springer-Verlag, 1997.